# CAN APPARENT BYSTANDERS DISTINCTIVELY SHAPE AN OUTCOME? GLOBAL SOUTH COUNTRIES AND GLOBAL CATASTROPHIC RISK-FOCUSED GOVERNANCE OF ARTIFICIAL INTELLIGENCE

Cecil Abungu,*1 Michelle Malonza*2 and Sumaya Nur Adan*3

## ABSTRACT

Increasingly, there is well-grounded concern that through perpetual scaling-up of computation power and data, current deep learning techniques will create highly capable artificial intelligence that could pursue goals in a manner that is not aligned with human values. In turn, such AI could have the potential of leading to a scenario in which there is serious global-scale damage to human wellbeing. Against this backdrop, a number of researchers and public policy professionals have been developing ideas about how to govern AI in a manner that reduces the chances that it could lead to a global catastrophe. The jurisdictional focus of a vast majority of their assessments so far has been the United States, China, and Europe. That preference seems to reveal an assumption underlying most of the work in this field: That global south countries can only have a marginal role in attempts to govern AI development from a global catastrophic risk-focused perspective. Our paper sets out to undermine this assumption. We argue that global south countries like India and Singapore (and specific coalitions) could in fact be fairly consequential in the global catastrophic risk-focused governance of AI. We support our position using 4 key claims. 3 are constructed out of the current ways in which advanced foundational AI models are built and used while one is constructed on the strategic roles that global south countries and coalitions have historically played in the design and use of multilateral rules and institutions. As each claim is elaborated, we also suggest some ways through which global south countries can play a positive role in designing, strengthening and operationalizing global catastrophic risk-focused AI governance.

* Authorship Note: The authors' names are listed in alphabetical order. All authors contributed equally to this work.

1 ILINA Program; University of Cambridge; Legal Priorities Project; Strathmore University. Corresponding email: cecil@ilinaprogram.org.

2 ILINA Program; AI Futures Fellowship. Corresponding email: michelle@ilinaprogram.org.

3 ILINA Program; University of Cambridge. Corresponding email: sumaya@ilinaprogram.org.

We are grateful to Amelia Midwa, Ben Garfinkel and Sienka Dounia, who helped us craft this paper. We are also grateful to the following people for helpful comments on the first draft of the paper: Bill Anderson-Samways, Haydn Belfield, Jai Vipra, Matthijs Maas, Oliver Guest, and Sebastien Krier.

# TABLE OF CONTENTS

*"Because global AI governance is only as good as the worst governed country, company, or technology, it must be watertight everywhere… A single loophole, weak link, or rogue defector will open the door to widespread leakage, bad actors, or a regulatory race to the bottom."*[4]

## 1. INTRODUCTION

There are plausible claims suggesting that when it emerges, very advanced artificial intelligence (AI) that does not pursue its goals in line with human values could lead to global catastrophe.[5] Some of the arguments depend on radical developments in AI capabilities while others rest on more "realistic" expectations given the progress that has already been made in capabilities.[6] The uninitiated might find the idea that AI could lead to global catastrophe a strange, sci-fi-esque flight of imagination. But at the core of any strand of this thought is the legitimate concern that: (i) it is still not possible to fully understand and guide what occurs when neural networks in deep learning models are engaged in decision-making,[7] (ii) AI development using deep learning continues apace[8] and (iii) the capabilities of AI systems are increasingly human-like.[9] Unless there is a breakthrough in understanding, alignment or governance, human beings could be leaving their fates to chance. Lately, there is more recognition of this risk at high echelons of government. For example, the White House's 2023 AI strategic plan notes that "long term risks remain, including the existential risk associated with the development of artificial general intelligence through self-modifying AI or other means."[10] Similarly, British Prime Minister Rishi Sunak has acknowledged the potential existential threat that "superintelligent" AI could carry[11] while the United Nations Secretary General has ranked the risk at par with that which nuclear war has.[12]

---

[4] Ian Bremmer and Mustafa Suleyman, 'The AI power paradox: Can states learn to govern artificial intelligence–Before it's too late?' Volume 102 Issue 5, Foreign Affairs, 2023, https://www.foreignaffairs.com/world/artificial-intelligence-power-paradox, page 39.

[5] Kaj Sotala and Roman Yampolskiy, 'Responses so catastrophic AGI risk: A survey,' Volume 90 Issue 1, Physica Scripta, 2015, https://iopscience.iop.org/article/10.1088/0031-8949/90/1/018001/pdf, pages 3-4; Karina Vold and Daniel Harris, 'How does artificial intelligence pose an existential risk,' in Carissa Veliz (editor), The Oxford Handbook of Digital Ethics, Oxford University Press, Oxford, 2021, pages 4-16.

[6] Paul Christiano, 'What failure looks like,' AI Alignment Forum, 17 March 2019, https://www.alignmentforum.org/posts/HBxe6wdjxK239zajf/what-failure-looks-like accessed on 8 August 2023.

[7] Richard Ngo et al, 'The alignment problem from a deep learning perspective,' arxiv, 2022, https://arxiv.org/pdf/2209.00626.pdf , pages 10-11; Ajeya Cotra, 'Why AI alignment could be hard with modern deep learning,' Cold Takes, 21 September 2021, https://www.cold-takes.com/why-ai-alignment-could-be-hard-with-modern-deep-learning/ accessed on 17 March 2023.

[8] Yoshua Bengio, 'Springtime for AI: The rise of deep learning,' Scientific American, 1 June 2016 https://www.scientificamerican.com/article/springtime-for-ai-the-rise-of-deep-learning/ accessed on 20 October 2023; Iqbal Sarker, 'Deep learning: A comprehensive overview on techniques, taxonomy, applications and research directions,' SN Computer Science, Volume 2 Issue 6, 2021, https://link.springer.com/article/10.1007/s42979-021-00815-1, pages 1, 3.

[9] Sebastian Bubeck et al, 'Sparks of artificial general intelligence: Early experiments with GPT-4,' arxiv, 2023, https://arxiv.org/pdf/2303.12712.pdf, pages 8-9,92; OpenAI, 'GPT-4 technical report,' arxiv, 2023, https://arxiv.org/pdf/2303.08774.pdf, page 14.

[10] Select Committee of National Intelligence of the National Science and Technology Council, 'National artificial intelligence research and development strategic plan 2023 update,' Executive Office of the President, May 2023, https://www.whitehouse.gov/wp-content/uploads/2023/05/National-Artificial-Intelligence-Research-and-Development-Strategic-Plan-2023-Update.pdf, page 17.

The situation has led to an exponential growth in efforts to find fitting and workable ideas on how to regulate AI in a manner that minimizes the chances of a global catastrophic risk occurring.[13] Embedded within most of these efforts is an overfocus on the countries where the leading AI companies are located and where the most significant investment is. Occasionally, there is also some attention on their peers.[14] While recent research has devoted more interest to the possible use of international legal instruments and institutions,[15] none of the work seems to explore how global south countries could shape any success or failure there. We think it is because global south countries seem to occupy a peripheral place in the building of advanced AI that many observers have—understandably—assumed that there is little to no role for them to play in influencing its creation, impact, and governance.

In this article we mainly argue that in the long run, global south countries such as India, Singapore, Brazil and others could be more influential than currently imagined in determining whether the world will experience AI which could potentially give rise to significant global catastrophe.[16] Like Joseph Carlsmith,[17] we think the highest risk will lie in AI systems that: (i) have advanced capabilities beyond that of "best humans" in high-level tasks; (ii) are able to make and implement plans and objectives based on real world models and (iii) have strategic awareness.[18] We define these kinds of AI systems as "highly capable AI" in this paper. Because it has already been done so comprehensively, we elected not to rehash the argument that Carlsmith and many others (cited in the preceding page) have made about how highly capable AI can potentially lead to a global catastrophe. Instead, we assume the reader will take such a perspective to be plausible.

---

[11] Alex Hern and Kiran Stacey, 'No 10 acknowledges existential risk of AI for the first time,' The Guardian, 25 May 2023, https://www.theguardian.com/technology/2023/may/25/no-10-acknowledges-existential-risk-ai-first-time-rishi-sunak accessed on 24 October 2023.

[12] Aljazeera, 'UN Chief Guterres backs proposal to form watchdog to monitor AI,' June 13 2023, https://www.aljazeera.com/news/2023/6/13/un-chief-guterres-backs-proposal-to-form-watchdog-to-monitor-ai#:~:text=And%20they%20are%20loudest%20from,must%20take%20those%20warnings%20seriously.%E2%80%9D accessed on 24 October 2023.

[13] There has been a surge in research, conference meetings and high-level discussions about the subject. For example, the UK now has a Frontier AI Taskforce (https://www.gov.uk/government/publications/frontier-ai-taskforce-first-progress-report/frontier-ai-taskforce-first-progress-report accessed on 2 November 2023) and has just hosted an international AI Safety Summit (https://www.gov.uk/government/topical-events/ai-safety-summit-2023 accessed on 2 November 2023).

[14] For example, consider the hiring round that Open Philanthropy – by far the leading grant maker in the space – has recently opened. For AI governance and policy positions, they make clear their preference for candidates with specialization in AI policy and policy development in the US, EU, UK and China. See Open Philanthropy, 'New roles on our global catastrophic risks team,' 26 September 2023, https://www.openphilanthropy.org/research/new-roles-on-our-gcr-team/ accessed on 2 October 2023.

[15] See for example Robert Trager et al, 'International governance of civilian AI: A jurisdiction certification approach,' arxiv, 2023, https://arxiv.org/pdf/2308.15514.pdf; Lewis Ho et al 'International institutions for advanced AI,' arxiv, 2023, https://arxiv.org/pdf/2307.04699.pdf; Matthijs Maas and Jose Jaime Villalobos, 'International AI institutions: A literature review of models, examples and proposals- AI Foundations Report 1,' Legal Priorities Project, September 2023, https://deliverypdf.ssrn.com/delivery.php?ID=725114106069109017077072079110070074039034001032090029097070105029094000077031109025033036016025122120037081106123103005105126039072071083031125030015118072107021101095015081110111021097081073100098117081086030122007030073127068030002001017026112080089&EXT=pdf&INDEX=TRUE.

[16] This study considers global catastrophic risk to be a risk that 'has the potential to inflict serious damage to human wellbeing on a global scale.' See Nick Bostrom and Milan Cirkovic, 'Introduction,' in Nick Bostrom and Milan Cirkovic (editors), Global Catastrophic Risks, Oxford University Press, New York, 2008, page 1.

[17] Joseph Carlsmith, 'Is power-seeking AI an existential risk?' arxiv, 2021, https://arxiv.org/pdf/2206.13353.pdf, pages 8-28.

Our supporting claims for the main argument are as follows. First, we propose that the domestic governance decisions of particular global south countries will automatically assume importance if: (i) Any of these countries builds on-premise supercomputers that can train highly capable AI; (ii) Cloud-based high performance computing continues to be used to build highly capable AI, and some of the crucial data centers are in any of these countries; (iii) Some AI developers continue to release powerful advanced AI models open-source and (iv) Reinforcement Learning from Human Feedback (RLHF) remains a crucial way to train models, and any of these countries continues to be an appealing site for RLHF. Second, we think some of these countries could play a significant role if international governance of AI development becomes a key way to prioritize safety. We arrive at that conclusion primarily through analogical reasoning around the roles that global south countries have previously played in the design and use of other multilateral rules and institutions.

Exposing the mistaken assumption at play is not merely a theoretical vanity project. This paper shows that the assumption leaves blind spots which global policy makers cannot afford to overlook given the stakes at play. Thus, the subordinate goal of the paper is to propose some specific regulatory and advocacy measures that particular actors in the consequential global south countries can take to reduce the chances that the world will experience a global catastrophe because of highly capable AI.

Of course, it should not be in contest that domestic governance of AI in global south countries will, in any event, be important if one is thinking about how to prepare communities there for the impact that transformative AI will have. Likewise, it will be important in figuring out how to prevent and mitigate certain kinds of AI harms faced by communities in those countries,[19] and could offer useful examples in governance for the rest of the world.[20] The locus of this paper is not in any of these points. Instead, our research is focused on AI development that could lead to models that have the potential to cause significant global catastrophes.

The paper will flow as follows: In part 2, we present the technical reasons why we think some global south countries could matter. By 'technical', we mean reasons related to the science of deep learning as it is so far. Immediately after each technical reason, we will sketch out some regulatory steps the countries in question can take in response. In part 3, we present the strategic reasons – essentially, the reasons united by a theme of the "soft" roles that global south countries usually play in making progress on difficult global challenges. Once again, each strategic reason will be followed by some ideas on practical ways to ensure these countries are not weak links for AI safety. Part 4 of the paper will conclude and then summarize some broad positions on the way forward. Due to the nature of our argument, we see this contribution as a unique addition to the ever-growing body of research on AI risk and AI governance.

---

[18] Joseph Carlsmith, 'Is power-seeking AI an existential risk?' arxiv, 2021, https://arxiv.org/pdf/2206.13353.pdf, pages 8-28.

[19] Chinmayi Arun, 'AI and the Global South: Designing for other worlds,' in Markus Dubber, Frank Pasquale and Sunit Das (editors), The Oxford Handbook of Ethics of AI, Oxford University Press, New York, 2020, page 589.

[20] Marie-Therese Png, 'At the tensions of south and north: Roles of Global South stakeholders in AI governance,' in Justin Bullock et al (editors), Oxford Handbook of AI Governance, Oxford University Press, London, 2022, page 27.

## 2. TECHNICAL REASONS WHY SOME GLOBAL SOUTH COUNTRIES COULD BE INFLUENTIAL IN THE PATH TO HIGHLY CAPABLE MISALIGNED AI

### 2.1 Some global south countries are potential zones for training highly capable AI

**The scaling hypothesis and the challenge of scaling computational power**

Research so far[21] seems to affirm the scaling hypothesis, which suggests that achieving highly capable AI might largely hinge on utilizing high levels of computational power and large amounts of training data.[22] Aggregating computational power in a way that allows highly complex tasks to be executed quickly and accurately is referred to as high-performance computing (HPC).[23] One straightforward way to do HPC is through using on-premise supercomputers.[24] However, the intricacies of chip production, which involve highly specialized manufacturing processes and advanced raw materials, make that a very capital-intensive endeavor.[25] Costs are amplified by the requirement for dust-free environments, precise machinery, and extensive research and development efforts.

---

[21] Jaime Sevilla et al, 'Compute trends across three eras of machine learning,' Institute of Electrical and Electronics Engineers, International Joint Conference on Neural Networks, Padua, pages 18-23 July 2022, https://doi.org/10.1109/IJCNN55064.2022.9891914, page 3; Ben Buchanan, 'The AI triad and what it means for national security strategy,' Center for Security and Emerging Technology, August 2020, https://cset.georgetown.edu/publication/the-ai-triad-and-what-it-means-for-national-security-strategy/, pages 7-9.

[22] Gwern Branwen, 'The Scaling Hypothesis,' Gwern.net, 28 May 2020, https://gwern.net/scaling-hypothesis accessed on 24 October 2023.

[23] Indeed, to spur the development of AI in their territories, countries like the US, China and EU countries are investing in supercomputers. See Katie Jones, 'Launching a new class of US supercomputing,' Energy.gov, 17 November 2022, https://www.energy.gov/science/articles/launching-new-class-us-supercomputing accessed on 13 August 2023; Xinhua News, 'China to build up supercomputing capacities,' 19 April 2023, http://english.news.cn/20230419/7849eacc49094dc58c5408dd36d7ff27/c.html accessed on 13 August 2023; European Commission, 'EU steps up investment in world class supercomputers for researchers and businesses,' 21 October 2020, https://digital-strategy.ec.europa.eu/en/news/eu-steps-investment-world-class-supercomputers-researchers-and-businesses accessed on 13 August 2023; Natasha Lomas, 'EU to let 'responsible' AI startups train models on its supercomputers,' Techcrunch, 13 September 2023, https://techcrunch.com/2023/09/13/eu-supercomputers-for-ai/ accessed on 13 August 2023.

[24] See Katie Jones, 'Launching a new class of US supercomputing,' Energy.gov, 17 November 2022, https://www.energy.gov/science/articles/launching-new-class-us-supercomputing accessed on 13 August 2023.

[25] Jai Vipra and Sarah West, 'Computational power and AI,' AI Now Institute, 27 September 2023, https://ainowinstitute.org/publication/policy/compute-and-ai accessed on 13 August 2023.

### A. Some global south countries could offer on-premise supercomputers that allow actors to train highly capable AI

One glance at India reveals what could happen there and in a range of other global south countries. India has taken significant steps towards the development of supercomputers in the last few decades. Through its National Supercomputing Mission, it has built and deployed 18 supercomputers.[26] Out of these, 4 feature in the 61st edition of Top500 Global Supercomputing List.[27] At the top is AIRAWAT, which was installed under the National Program on AI by the Government of India at the Centre for Development of Advanced Computing (C-DAC).[28] AIRAWAT was ranked 75th globally at the International Supercomputing Conference 2023[29] and it holds the same position on the Top500 list.[30] Several other Indian supercomputers are not far behind.

It would not be a stretch to claim that the most capable of these supercomputers have primarily been built to train AI models. Indeed, it is on record that they are part of an elaborate, multi-pronged national strategy to make India a global leader in AI by 2030.[31] This strategy is guided by the National Strategy for Artificial Intelligence and the National Supercomputing Mission.[32] In the private sphere, Nvidia and Reliance Industries of India have recently announced a partnership to build AI supercomputers. The plan is to build AI infrastructure that is more powerful than the fastest supercomputer (AIRAWAT). Their collaboration is focused on building AI computing resources that will be used to develop homegrown advanced models. The AI infrastructure will be hosted in AI-ready computing data centers maintained by Reliance Industries.[33]

---

[26] Amit Chaturvedi, 'India to get 9 more supercomputers: list of fastest machines in the country,' NDTV, 17 August 2023, https://www.ndtv.com/india-news/india-to-get-9-more-supercomputers-list-of-fastest-machines-in-country-4304973 accessed on 18 October 2023; FP Explainers, 'India to get nine more supercomputers: which are the fastest in the country?' 17 August 2023, https://www.firstpost.com/explainers/india-to-get-nine-more-supercomputers-which-are-the-fastest-in-the-country-13009072.html/amp accessed on 18 October 2023.

[27] Times of India, '9 more supercomputers to be added under National Supercomputing Mission,' 17 August 2023, https://timesofindia.indiatimes.com/gadgets-news/9-more-supercomputers-to-be-added-under-national-supercomputing-mission-list-of-four-current-fastest-machines-in-india/articleshow/102786359.cms accessed on 18 October 2023; TOP500, 'TOP500 list,' June 2023, https://www.top500.org/lists/top500/list/2023/06/ accessed on 18 October 2023.

[28] Times of India, '9 more supercomputers to be added under National Supercomputing Mission,' 17 August 2023, https://timesofindia.indiatimes.com/gadgets-news/9-more-supercomputers-to-be-added-under-national-supercomputing-mission-list-of-four-current-fastest-machines-in-india/articleshow/102786359.cms accessed on 18 October 2023.

[29] Times of India, 'India's AI Supercomputer 'AIRAWAT' makes it to the list of world's 100 most powerful,' 17 August 2023, https://timesofindia.indiatimes.com/gadgets-news/indias-ai-supercomputer-airawat-makes-it-to-the-list-of-worlds-100-most-powerful/amp_articleshow/100487763.cms accessed on 18 October 2023.

[30] TOP500, 'TOP500 list,' June 2023, https://www.top500.org/lists/top500/list/2023/06/ accessed on 18 October 2023.

[31] India AI 2023, 'Expert Group report to the Ministry of Electronics and Information Technology- First edition,' October 2023, https://www.meity.gov.in/writereaddata/files/IndiaAI-Expert-Group-Report-First-Edition.pdf, pages 131-132.

[32] India AI 2023, 'Expert Group report to the Ministry of Electronics and Information Technology- First edition,' October 2023, https://www.meity.gov.in/writereaddata/files/IndiaAI-Expert-Group-Report-First-Edition.pdf, pages 131-132.

We are convinced that this example, coupled with (i) the often underrecognized power of nationalistic motivation and (ii) the fact that global south countries like India are unlikely to be banned from accessing the most powerful computer chips available, shows that it is fairly likely that some global south countries could have supercomputing capacity to train highly capable AI. The likelihood will probably be accelerated if a narrative that having home-built highly capable AI is crucial to a country's development (further explored in part 3 of this paper) takes root.

### B. Colocation data centers in some global south countries could be used in the training of highly capable AI

#### (i) The rise of cloud-based high-performance computing

Apart from using on-premise supercomputers, HPC can also be done over the internet through cloud computing. Essentially, this would mean that computing tasks are executed over a global network of servers on the internet.[34] This has become a popular way to access scaled computational power because it allows for cost-saving and offers flexibility.[35] Consequently, leading AI companies like OpenAI[36], Anthropic[37] and Google DeepMind[38] have established strategic partnerships with prominent cloud service providers such as Microsoft, Google and Amazon. There are already real results. GPT-4, for example, was trained on cloud-based HPC that Microsoft specially built for OpenAI.[39]

#### (ii) The place of data centers in cloud-based HPC

Data centers serve as a linchpin between the realms of HPC and cloud services, featuring as indispensable assets for compute-intensive tasks and storage requirements.[40] On one hand, they directly facilitate HPC by providing the robust physical infrastructure—equipped with high-speed networking, formidable power, adept cooling systems, and stringent security protocols—necessary to manage and process complex, voluminous data sets and workloads with superior efficacy.[41]

---

[33] Times of India, 'Reliance-NVIDIA to build AI supercomputers in India,' 11 September 2023, https://timesofindia.indiatimes.com/business/india-business/reliance-nvidia-to-build-ai-supercomputers-in-india/articleshow/103506497.cms accessed on 17 October 2023.

[34] David Linthicum, 'The cloud as a supercomputer,' InfoWorld, 30 November 2021, https://www.infoworld.com/article/3642848/the-cloud-as-supercomputer.html accessed on 25 October 2023.

[35] Rescale, 'What is cloud high performance computing?' https://rescale.com/cloud-hpc/ accessed on 25 October 2023.

[36] Microsoft Corporate Blogs, 'Microsoft and OpenAI extend partnership,' Official Microsoft Blogs, 23 January 2023, https://blogs.microsoft.com/blog/2023/01/23/microsoftandopenaiextendpartnership/ accessed on 25 October 2023.

[37] Anthropic, 'Expanding access to safer AI with Amazon,' 25 September 2023, https://www.anthropic.com/index/anthropic-amazon accessed on 25 October 2023.

[38] Samuel Gibbs, 'Google buys UK artificial intelligence startup Deepmind for £400,' The Guardian, 27 January 2014, https://www.theguardian.com/technology/2014/jan/27/google-acquires-uk-artificial-intelligence-startup-deepmind accessed on 25 October 2023.

[39] John Roach, 'How Microsoft's bet on Azure unlocked an AI revolution,' Microsoft News, 13 March 2023, https://news.microsoft.com/source/features/ai/how-microsofts-bet-on-azure-unlocked-an-ai-revolution/ accessed on 25 October 2023.

[40] Sebastian Moss, 'Generative AI & the future of data centers: Part VII- The data centers,' Data Center Dynamics, 13 July 2023, https://www.datacenterdynamics.com/en/analysis/generative-ai-the-future-of-data-centers-part-vii-the-data-centers/ accessed on 25 October 2023.

On the other hand, data centers are crucial fortresses for cloud computing providers, hosting the physical hardware (such as servers and storage devices) that underpin the virtualized computing resources offered over the internet.[42] Merging these dynamics, cloud providers enable organizations to harness the promise of HPC without having to make hefty investments in dedicated infrastructure. They allow scalable, on-demand access to colossal computational power and storage, thereby empowering entities to efficiently process and analyze extensive datasets in a flexible and economically savvy manner. Additionally, prominent cloud computing providers are offering highly competitive pricing to AI companies, even if it means operating without immediate profitability during the initial stages.[43]

Central to our analysis is the emergence and increasing dominance of colocation data centers. These multi-tenant facilities have proven indispensable for actors not looking to build their own data centers.[44] They allow several actors to lease capacity from colocation providers, who manage the supporting infrastructure, including power supply, cooling, and connectivity. Although there are tens of thousands of data centers in the world, researchers have argued that only about 325 to 1400 of them are currently large enough (10+ megawatts) to host AI supercomputers.[45]

### (iii) Global south countries could host crucial colocation data centers

The proposition that global south countries could host data centers is not novel because data centers already exist in some of these countries. A 2021 US International Trade Commission Report indicates that Brazil and India are already home to 4% of all of the world's data centers.[46] Further evidence can be found in the fact that in Southeast Asia, countries including Singapore, Indonesia, Thailand and Malaysia are sites for data centers run by the likes of Amazon, Equinix, Meta, Google, Microsoft and Equinix.[47] Even in Africa,

---

[41] Konstantin Pilz and Lennart Heim, 'Compute at scale: A broad investigation into the data center industry,' Konstantin F. Pilz, July 2023, https://www.konstantinpilz.com/data-centers/report, page 11.

[42] Sebastian Moss, 'Generative AI & the future of data centers: Part VII- The data centers,' Data Center Dynamics, 13 July 2023, https://www.datacenterdynamics.com/en/analysis/generative-ai-the-future-of-data-centers-part-vii-the-data-centers/ accessed on 25 October 2023.

[43] Konstantin Pilz and Lennart Heim, 'Compute at scale: A broad investigation into the data center industry,' Konstantin F. Pilz, July 2023, https://www.konstantinpilz.com/data-centers/report, page 27.

[44] Konstantin Pilz and Lennart Heim, 'Compute at scale: A broad investigation into the data center industry,' Konstantin F. Pilz, July 2023, https://www.konstantinpilz.com/data-centers/report, page 17.

[45] Konstantin Pilz and Lennart Heim, 'Compute at scale: A broad investigation into the data center industry,' Konstantin F. Pilz, July 2023, https://www.konstantinpilz.com/data-centers/report, page 2.

[46] Brian Daigle, 'Data centers around the world: A quick look,' United States International Trade Commission Executive Briefings on Trade, May 2021,
https://www.usitc.gov/publications/332/executive_briefings/ebot_data_centers_around_the_world.pdf,
page 2.

[47] CIO, 'Why Southeast Asia is the new Mecca for data centers,' 19 September 2019,
https://www.cio.com/article/222083/why-are-businesses-relocating-data-centres-to-southeast-asia.html
accessed on 25 October 2023.

South Africa plays host to data centers owned by Amazon[48] and Microsoft.[49] Even if we go by the argument that you currently need a minimum 10+ megawatts (MW) data center to host a compute cluster that can train the most capable AI models,[50] there is credible reason to believe that many data centers in global south countries (owned by the key cloud-based HPC providers) meet this minimum. For example, the Amazon data center in South Africa is served by a dedicated 10 MW solar farm.[51] In Singapore, Microsoft also has a 60 MW solar project to power its data centers,[52] and has been assured a share of 80 MW to develop new ones.[53] More pointedly, Amazon and Microsoft have each announced plans to build data centers with over 1 gigawatt capacity in India over the next 10 years.[54]

We may soon see more large data centers in global south countries. A study on the location of data centers owned and operated by Google in 2022[55] implied that when making decisions about where to locate them, the following are preferred: cold areas, areas that support renewable energy production and physically large areas (presumably to host the data rooms and to produce renewable energy). In comparison to global north countries, global south countries generally have more physical space available at lower costs. Generally, their governments are also a lot more desperate to attract investment, meaning they are more likely to create enormous incentives for cloud computing providers[56] (for example land curve outs and tax incentives). Finally, many global south

---

[48] Rachel England, 'Amazon opens its first cloud data center in Africa,' Engadget, 22 April 2020, https://www.engadget.com/amazon-opens-its-first-cloud-data-center-in-africa-101534093.html accessed on 25 October 2023.

[49] Microsoft Azure, 'Microsoft opens first data centers in Africa with general availability of Microsoft Azure,' 6 March 2019, https://azure.microsoft.com/en-us/blog/microsoft-opens-first-datacenters-in-africa-with-general-availability-of-microsoft-azure/ accessed on 25 October 2023.

[50] Konstantin Pilz and Lennart Heim, 'Compute at scale: A broad investigation into the data center industry,' Konstantin F. Pilz, July 2023, https://www.konstantinpilz.com/data-centers/report, page 20.

[51] Amazon Staff, 'Amazon's first South African solar plant delivers energy and opportunity,' 22 February 2022: https://www.aboutamazon.com/news/aws/amazons-first-south-african-solar-plant-delivers-energy-and-opportunity accessed on 25 October 2023.

[52] Microsoft News Center, 'Microsoft and Sunseap sign agreement on largest ever solar project in Singapore,' 28 February 2018, https://news.microsoft.com/2018/02/28/microsoft-and-sunseap-sign-agreement-on-largest-ever-solar-project-in-singapore/ accessed on 20 October 2023; Asha Barbaschow, 'Microsoft to power Singapore datacenter services with rooftop solar,' Zdnet, 28 February 2018, https://www.zdnet.com/article/microsoft-to-power-singapore-datacentre-services-with-rooftop-solar/ accessed on 20 October 2023.

[53] Rebecca Uffindell, 'Singapore awards four data center operators with 80MW of new capacity,' Techerati, 17 July 2023, https://www.techerati.com/news-hub/singapore-awards-four-data-centre-operators-80mw-new-capacity-airtrunk-bytedance-equinix-gds-microsoft-moratorium/ accessed on 21 October 2023; Dan Swinhoe, 'Singapore selects Equinix, Microsoft, AirTrunk and GDS for 80MW data center trial,' Data Center Dynamics, 14 July 2023, https://www.datacenterdynamics.com/en/news/singapore-selects-equinix-microsoft-airtrunk-and-gds-for-80mw-data-center-trial/ accessed on 21 October 2023.

[54] Pratap Singh, 'AWS, Microsoft may take the wind out of data center firms' sails with in-house capacity,' The Ken, 22 December 2022, https://the-ken.com/story/aws-microsoft-may-take-the-wind-out-of-data-centre-firms-sails-with-in-house-capacity/ accessed on 25 October 2023.

[55] Patricia Arroba et al, 'Sustainable edge computing: Challenges and future directions,' arxiv, 2023, https://arxiv.org/pdf/2304.04450.pdf, pages 1-2.

[56] Instructively, India and South Africa are two of the three top countries that have received the most foreign direct investment in data centers since 2019. See Sebastian Shehadi, 'India is the top destination for foreign direct investment in data centers,' Investment Monitor, 27 September 2023,

countries rank quite competitively in fiber connectivity,[57] which is now considered critical for facilitating cloud computing.[58]

### (iv) Does the distance of global south countries mean the data centers in their countries are unlikely to be used in the training of highly capable AI models?

Evidence so far seems to show that there is a preference for executing large-scale AI training runs in locations close to the AI companies' principal physical presence. For example, it appears that OpenAI used a supercomputer in Ohio (USA) to train GPT-3[59] while GPT-4 was trained in Iowa (USA).[60] The reason behind this is not entirely clear. Some have claimed it has to do with the costs of labor and electricity, but others have claimed it has to do with national security.[61] Still, others have claimed that the decision has something to do with latency. One report notes that "experts have said it can make sense to pretrain an AI model in a single location because of the large amounts of data that need to be transferred between computing cores."[62]

That said, there are reasons to believe that in the future, highly capable AI built by leading AI companies may be trained in several locations further away from their headquarters. And some of these may be cloud based HPC data centers in global south countries. When running Internet of Things (IoT) applications, 'edge computing' (where the concerned colocation data center is as physically close as possible) is preferable because it ensures timely and accurate responses.[63] But training AI models is not the same as running AI applications. Running AI applications smoothly requires very frequent exchange of data between users and host data centers. On the other hand, training AI does not necessarily require that. In line with this perspective, some analysts have noted that:

> "Data centers for training AI models are less reliant on latency so proximity to the end user is not a key consideration … Access to robust power grids, dense fiber optic networks and cloud region internet exchange points will ensure fast and stable data transmission."[64]

Even if this position is incorrect and latency is in fact a significant concern, there are ever-advancing methods (for example, edges and nodes) to make latency very low despite the distance.[65] Additionally, as discussed in the part preceding this one, fiber connection in some global south countries is excellent. Finally, the uniqueness of infrastructures in several global south countries means that their networks are often not as congested as those in global north countries.[66]

All this of course does not take into account the fact that data protection regulation may well force these AI companies to start doing some training runs outside their home countries. Given many countries' enthusiasm about data sovereignty, we do not think it would be entirely unexpected for these companies to be forced to train advanced AI models in some global south countries (for example, India) whose valuable sizes of data cannot be ignored. This is especially true given the endless search for more data to train on.

### C. How global south countries can reduce the chances that highly capable misaligned AI will be trained in their jurisdictions

In response to the possibility that certain actors could have access to the computing power required to build highly capable AI, we think the global south countries like India should consider 2 types of regulation. The first is regulation requiring actors (who intend to train large AI models) seeking access to government-sponsored on-premise supercomputers like AIRAWAT to comply with certain rules. The rules we foresee here are the ones designed to reduce the chances that highly capable misaligned AI will emerge. Some include detailed reporting of large training runs[67] and mandatory pauses in large training runs when the models being trained reveal significant capabilities and

---

[64] Daniel Thorpe, '4 strategic location factors for AI training data centers,' JLL, 27 September 2023, https://www.jll.co.uk/en/trends-and-insights/research/talking-points/four-strategic-location-factors-for-ai-training-data-centres#:~:text=Countries%20with%20a%20thriving%20research,collaboration%20between%20academia%20and%20industry accessed on 25 October 2023.

[65] Nir Barazida, 'Distributed training of deep learning models: handling stragglers and latency in synchronous training,' Towards Data Science, 8 March 2022, https://towardsdatascience.com/stragglers-and-latency-in-synchronous-distributed-training-of-deep-learning-models-43783b0266d9 accessed on 25 October 2023; Haochen Hua et al, 'Edge computing with artificial intelligence: A machine learning perspective,' Volume 5 Issue 9, ACM Journals Computing Surveys, 2023, https://dl.acm.org/doi/pdf/10.1145/3555802, page 4-7.

[66] This conclusion is reached through a consideration of the number of connected users, packets being exchanged, requests being made, and upgrades being made to their infrastructures.

[67] Yonadav Shavit, 'What does it take to catch a Chinchilla? Verifying rules on large-scale neural network training via compute monitoring,' arxiv, 2023, https://arxiv.org/pdf/2303.11341.pdf, pages 4-12.

misalignment.⁶⁸ In a way, this approach would mirror that which the EU has instituted for access to its on-premise supercomputing facilities.⁶⁹

The second type of regulation should be that which requires private actors who manage supercomputers (for example whatever is likely to come out of the Nvidia-Reliance agreement) to monitor and report large training runs, and perhaps to comply with certain "Know Your Customer" rules⁷⁰ if they rent out their supercomputers out for use.

It may seem difficult to see how global south countries could respond to the possibility that leading AI companies may conduct large training runs using colocation data centers in their jurisdictions. Since these AI companies are likely to remain primarily headquartered in the US, American regulation is probably the key way to keep them on a tight leash. However, global south countries can still be useful. The governments of these countries could set high regulatory penalties for reporting failures⁷¹ and conduct random inspection tours of the data centers in question. Additionally, these countries could make it an official policy to cooperate closely with any American or international regulatory body that is charged with ensuring that the rules for large scale AI training are being followed.

## 2.2 Some global south countries could matter if Reinforcement Learning from Human Feedback remains important in building highly capable AI

### A. Understanding Reinforcement Learning from Human Feedback (RLHF)

Reinforcement learning from human feedback (RLHF) is essentially the alteration of model behavior through the intervention of human beings in the process of training. According to Lambert and Huyen, it occurs in 3 stages: (i) creation of a base model, (ii) preference modelling and (iii) integration of the preference model with the base model (the actual reinforcement learning).⁷² So far, RLHF has earned its stripes through the advancement of large language models (LLM). As a result, understanding it requires understanding its use in developing LLMs.

The creation of a very capable and helpful base model is the starting point of RLHF. After creation of the base model, preference modelling begins—and human labelers play a key role here. The base model is fed prompts and used to generate a large batch of responses. Humans are then called to determine the better responses to the prompts. A prompt and its chosen response are then labelled as a preference pair. More specifically, the harmful prompt is paired with a response that 'eliminates substantially answering' of the harmful prompt, in the lines of "as a language model I don't want to answer that".[73] The final step is integrating the preference data into the base model (extracting information from the preference model).[74]

---

⁶⁸ Markus Anderljung et al, 'Frontier AI regulation: Managing emerging risks to public safety,' arxiv, 2023, https://arxiv.org/pdf/2307.03718.pdf, page 28.

⁶⁹ Natasha Lomas, 'EU to let 'responsible' AI startups train models on its supercomputers,' Techcrunch, 13 September 2023, https://techcrunch.com/2023/09/13/eu-supercomputers-for-ai/ accessed on 13 August 2023.

⁷⁰ Janet Egan and Lennart Heim, 'Oversight for Frontier AI through a Know-Your-Customer Scheme for Compute Providers' arxiv, 2023, https://cdn.governance.ai/Oversight_for_Frontier_AI_through_a_KYC_Scheme_for_Compute_Providers.pdf, page 2-3.

⁷¹ Yonadav Shavit, 'What does it take to catch a Chinchilla? Verifying rules on large-scale neural network training via compute monitoring,' arxiv, 2023, https://arxiv.org/pdf/2303.11341.pdf, page 4.

⁷² Nathan Lambert, 'How RLHF actually works,' Interconnects, 21 June 2023, https://www.interconnects.ai/p/how-rlhf-works accessed on 14 July 2023; Chip Huyen, 'RLHF: Reinforcement Learning from Human Feedback,' 2 May 2023, https://huyenchip.com/2023/05/02/rlhf.html accessed on 19 July 2023.

## B. The place of RLHF in the development of highly capable AI

Although it is still imperfect,[75] experiments have shown that where RLHF has been used in the training of LLMs, the resulting models show impressive improvements in helpfulness, honesty, and harmlessness.[76]

Researchers have—for example—found that AI agents can be trained to be helpful, honest, and harmless through preference modelling and RLHF.[77] The researchers trained a preference model on 2 datasets: one focused on helpfulness and the other focused on harmfulness. The preference models trained on a mixture of both datasets behaved helpfully when appropriate and politely declined harmful requests. Reinforcement learning then followed where the base model was optimized using the preference model scores as rewards. The resultant model performed better than the base model on virtually all evaluations hence confirming that alignment training does not necessarily compromise AI capabilities (typically referred to as "AI tax").[78] So far, RLHF is therefore arguably the best discovered way to "align" models.

It is no wonder then that OpenAI[79] and Google[80] are currently using RLHF to fine tune their LLMs. Yet one might still wonder: Does RLHF's preeminence in the training of LLMs say anything about whether it might be useful in the training of highly capable AI in the long-term future? Although there is significant skepticism that RLHF will play a key role in aligning such AI,[81] we think there are 3 ways RLHF may still matter a lot. First, unless a better method is devised, RLHF is likely to continue being a principal method through which advanced AI systems are "aligned",[82] and one of those advanced AI systems could just end up being highly capable AI (as we define it).[83] Second, even if the first point is unconvincing, RLHF's measured success gives cover for AI companies to present their advanced AI models as "safe enough" – and thus argue that development of highly capable AI should not be limited.[84] Third, it may be useful for the "iterative experimentation" that gives an actor the confidence to build highly capable AI.[85]

---

[73] Nathan Lambert, 'How RLHF actually works,' Interconnects, 21 June 2023, https://www.interconnects.ai/p/how-rlhf-works accessed on 14 July 2023.

[74] Nathan Lambert, 'How RLHF actually works,' Interconnects, 21 June 2023, https://www.interconnects.ai/p/how-rlhf-works accessed on 14 July 2023; Long Ouyang et al, 'Training language models to follow Instructions with human feedback, arxiv, 2022, https://arxiv.org/pdf/2203.02155.pdf, pages 1,9; Yuntao Bai et al, 'Training a helpful and harmless assistant with reinforcement learning from human feedback,' arxiv, 2022, https://arxiv.org/pdf/2204.05862.pdf, pages 16-17; Chip Huyen, 'RLHF: Reinforcement Learning from Human Feedback,' 2 May 2023, https://huyenchip.com/2023/05/02/rlhf.html accessed on 19 July 2023.

[75] Long Ouyang et al, 'Training language models to follow Instructions with human feedback, arxiv, 2022, https://arxiv.org/pdf/2203.02155.pdf, pages 3-4; Yufei Wang et al, 'Aligning large language models with human: A survey,' arxiv, 2023, https://arxiv.org/pdf/2307.12966.pdf, pages 12-14.

[76] Long Ouyang et al, 'Training language models to follow Instructions with human feedback, arxiv, 2022, https://arxiv.org/pdf/2203.02155.pdf, pages 3-4; Yuntao Bai et al, 'Training a helpful and harmless assistant with reinforcement learning from human feedback,' arxiv, 2022, https://arxiv.org/pdf/2204.05862.pdf, pages 22-23.

[77] Yuntao Bai et al, 'Training a helpful and harmless assistant with reinforcement learning from human feedback,' arxiv, 2022, https://arxiv.org/pdf/2204.05862.pdf, pages 22-23.

[78] Yuntao Bai et al, 'Training a helpful and harmless assistant with reinforcement learning from human feedback,' arxiv, 2022, https://arxiv.org/pdf/2204.05862.pdf, pages 4-5, 13-23.

[79] Jan Leike, John Schulman and Jeffrey Wu, 'Our Approach to alignment research,' OpenAI, 24 August 2022, https://openai.com/blog/our-approach-to-alignment-research accessed on 5 August 2023.

[80] James Manyika, 'An overview of Bard: An early experiment with generative AI,' Google AI, 10 April 2023, https://ai.google/static/documents/google-about-bard.pdf accessed on 5 August 2023, page 3.

That said, there is a chance that in the long run RLHF may be displaced as a leading alignment method. RLHF is fundamentally outcome supervision. This is an important fact because recent research suggests that its opposite – process supervision – could be more effective at aligning AI than outcome supervision.[86] Additionally, researchers are working on other methods of alignment that do not involve humans (for example, reinforcement learning from AI feedback[87] and reinforcement learning from contrast distillation[88]). If any of these or other alternatives[89] is eventually adopted as standard-setting, then perhaps our point here will dissolve. However, it seems more likely to us that any preferred method that emerges will combine both process supervision and outcome supervision with humans still playing a major role (the way Anthropic's 'Constitutional AI' method uses AI feedback to limit model harmfulness and human feedback to increase model helpfulness[90]).

---

[81] Leopold Aschenbrenner, 'Nobody's on the ball on AGI alignment,' Effective Altruism Forum, 29 March 2023, https://forum.effectivealtruism.org/posts/5LNxeWFdoynvgZeik/nobody-s-on-the-ball-on-agi-alignment accessed on 26 October 2023.

[82] Yufei Wang et al, 'Aligning large language models with human: A survey,' arxiv, 2023, https://arxiv.org/pdf/2307.12966.pdf, pages 13.

[83] Bubeck S et al, 'Sparks of Artificial General Intelligence: Early Experiments with GPT-4,' arxiv, 2023, https://arxiv.org/pdf/2303.12712.pdf, page 92.

[84] Jan Leike, John Schulman and Jeffrey Wu, 'Our Approach to alignment research,' OpenAI, 24 August 2022, https://openai.com/blog/our-approach-to-alignment-research accessed on 5 August 2023; James Manyika, 'An overview of Bard: An early experiment with generative AI,' Google AI, 10 April 2023, https://ai.google/static/documents/google-about-bard.pdf accessed on 5 August 2023, page 3.

[85] Leopold Aschenbrenner, 'Nobody's on the ball on AGI alignment,' Effective Altruism Forum, 29 March 2023, https://forum.effectivealtruism.org/posts/5LNxeWFdoynvgZeik/nobody-s-on-the-ball-on-agi-alignment accessed on 26 October 2023.

[86] Hunter Lightman et al, 'Let's verify step by step,' arxiv, 2023, https://arxiv.org/pdf/2305.20050.pdf, page 11.

[87] Yuntao Bai et al, 'Constitutional AI: Harmlessness from AI feedback,' arxiv, 2022, https://arxiv.org/pdf/2212.08073.pdf.

[88] Kevin Yang et al, 'Reinforcement learning from contrast distillation for language model alignment,' arxiv, 2023, https://arxiv.org/pdf/2307.12950.pdf.

[89] For example, one could imagine the new 'superalignment' project at OpenAI yielding some "scientific and technical breakthroughs." See Jan Leike and Ilya Sutskever, 'Introducing superalignment,' OpenAI, 5 July 2023 https://openai.com/blog/introducing-superalignment accessed on 26 October 2023.

[90] Yuntao Bai et al, 'Constitutional AI: Harmlessness from AI Feedback,' arxiv, 2022, https://arxiv.org/pdf/2212.08073.pdf, pages 8-14.

## C. The relevance of global south countries in RLHF

As demonstrated earlier, data labelling is a critical part of RLHF since RLHF only becomes possible after data labelling. At the moment, global south countries provide what has been described as a "vast tasker underclass" that allows for cheap data labelling during AI development.[91] Even though AI companies are often opaque about who they hire to label data, some reports have shown that workers in global south countries are a very important part of the matrix. For example, it is on record that Sama, a San Francisco-based firm that employs workers in Kenya, Uganda and India, was contracted by OpenAI to provide labelling services for training models to limit biased and harmful outputs.[92] No wonder analysts like Kori Hale have referred to data labelers in Africa as the "construction workers" of the digital world.[93]

There is an obvious reason for these AI companies to source labor from African countries. Wages are low and it generally saves them a lot of money. For example, Sama is said to have hired data labelers in Kenya at around 13-16 USD per day for repetitive labelling, which accounts for a lot of the time spent in AI development. It also seems difficult for these companies to find people in global north countries to perform the boring, monotonous, and never-ending task of data labelling at a low cost.[94]

There is some evidence that RLHF labelling work is also done by workers from other parts of the world as well. To do one important RLHF study, for example, OpenAI sourced its 40 labelers from Upwork and Scale AI. These 2 platforms have freelancers from all over the world. Their nationality is unclear: they simply had to pass a screening test.[95] For the Anthropic study on RLHF, the labelers were master-qualified US-based MTurk workers (MTurk is a crowdsourcing marketplace that makes it easier for businesses to outsource tasks that are performed virtually). Anthropic is also on record as having recruited some labelers from Upwork.[96] Nevertheless, it still appears to be the case that significant amounts of human RLHF work goes on in global south countries. Indeed, Lambert has acknowledged this and proposed that the difference in the RLHF work going on in different parts of the world may be related to the level of education required for a task (with more sophisticated tasks being done in global north countries).[97]

---

[91] Josh Dzieza, 'AI is a lot of work,' The Verge, 20 June 2023, https://www.theverge.com/features/23764584/ai-artificial-intelligence-data-notation-labor-scale-surge-remotasks-openai-chatbots accessed on 21 July 2023.

[92] Billy Perrigo, 'OpenAI used Kenyan workers on less than $2 per hour to make ChatGPT less toxic,' Time, 18 January 2023, https://time.com/6247678/openai-chatgpt-kenya-workers/ accessed on 21 July 2023.

[93] Kori Hale, 'Google and Microsoft are banking on Africa's AI labeling workforce,' Forbes, 28 May 2019, https://www.forbes.com/sites/korihale/2019/05/28/google-microsoft-banking-on-africas-ai-labeling-workforce/?sh=7a3f77e541c4 accessed on 21 July 2023.

[94] Kori Hale, 'Google and Microsoft are banking on Africa's AI labeling workforce,' Forbes, 28 May 2019, https://www.forbes.com/sites/korihale/2019/05/28/google-microsoft-banking-on-africas-ai-labeling-workforce/?sh=7a3f77e541c4 accessed on 21 July 2023.

[95] Long Ouyang et al, 'Training language models to follow Instructions with human feedback, arxiv, 2022, https://arxiv.org/pdf/2203.02155.pdf, page 36; Cade Metz, 'The secret ingredient of ChatGPT is human advice,' The New York Times, 25 September 2023, https://www.nytimes.com/2023/09/25/technology/chatgpt-rlhf-human-tutors.html?smid=nytcore-android-share accessed on 26 October 2023.

[96] Yuntao Bai et al, 'Training and Helpful and Harmless Assistant with Reinforcement Learning from Human Feedback,' arxiv, 2022, https://arxiv.org/pdf/2204.05862.pdf, pages 63-64.

### D. How global south countries can use the RLHF roles played by actors in their jurisdictions to reduce the chances that highly capable misaligned AI will emerge

First, we think these global south countries can strengthen labor regulations governing data labelling work. For example, they could raise the minimum wage for this work and require additional psychosocial support to be provided by the companies being contracted by the likes of OpenAI. Assuming that AI companies would not want to develop and deploy severely misaligned models, this step would almost certainly make RLHF more expensive, which would in turn raise barriers to building highly capable AI. Of course, such a measure would also help to address the human rights concerns that many have raised.[98]

Second, these countries could also raise taxes on the particular industry. Practically speaking, that may mean that the companies being contracted by OpenAI to help with the labelling pay more in taxes. Again, this would serve to make RLHF more expensive and therefore limit rational actors' ability to build highly capable AI. One could argue that no global south country can plausibly take this particular step given the economic downside it will have. In response, we suggest that the countries could be incentivized to take the step through aid and loan conditions.

While our suggested interventions may make it more expensive to carry out RLHF, it is possible that the leading AI companies will have enough resources[99] to cover any higher costs. In such an instance, our proposals may appear anodyne at best. However, we believe this will depend on other factors (for example, just how expensive the method becomes as well as the level of other costs that the companies have) that we did not set out to explore in this study.

## 2.3 Some global south countries have actors who could fine tune open-source foundational models and that could lead to highly capable misaligned AI

Leading AI companies and researchers have taken different positions as to whether the code and weights of the most advanced AI models ought to be made open source. Naturally, this dispute has played out in the realm of the most advanced models available right now: LLMs. Some have argued that the open-sourcing levels the playing field for other developers, which allows for faster development and innovation and provides an opportunity for people to scrutinize the software carefully.[100] Others have claimed that open sourcing such models is dangerous and irresponsible because it creates an opportunity for people to fine tune the model and build a less aligned one.[101]

---

[97] https://x.com/natolambert/status/1660060619331231744?s=46&t=bU0p5ELIRKO5LtijMLrFrA accessed on 2 November 2023.

[98] Karen Hao and Deepa Seetharaman, 'Cleaning up ChatGPT takes a heavy toll on human workers,' Wall Street Journal, 24 July 2023, https://www.wsj.com/articles/chatgpt-openai-content-abusive-sexually-explicit-harassment-kenya-workers-on-human-workers-cf191483 accessed on 26 October 2023.

[99] The leading AI companies continue to grow significantly in market capitalization. See for example Brian Fung, 'Amazon invests up to $4 billion in Anthropic AI in exchange for minority stake and further AWS integration,' CNN, September 25 2023, https://edition.cnn.com/2023/09/25/tech/amazon-invests-anthropic-ai/index.html accessed on 26 October 2023; Chris Metinko, 'Excited and scary: After sizzling first half, don't expect AI investment to cool off,' Crunchbase news, 6 July 2023, https://news.crunchbase.com/ai-robotics/q2-2023-open-ai-investment-coreweave-charts/ accessed on 26 October 2023.

### A. Understanding fine tuning of open-source frontier models

To understand what fine tuning an open-source model could possibly lead to, we think it is worth focusing on how fine tuning in LLMs works, since LLMs are the most capable models we have at the moment. Imitation learning is one of the most effective strategies to fine tune new LLMs from the outputs of already existing LLMs. It works through the process of knowledge distillation, which is the use of a large fully trained neural network as a training signal for another small neural network. The method effectively combines the usual training data of the small neural network with the output of the larger, more powerful neural network. There are 2 types of model imitation: local imitation and broad imitation. In local imitation, the smaller 'student' model learns to imitate a larger model's behavior in completing a *specific* task while in broad imitation it learns to broadly imitate a larger model's behavior over a *variety* of tasks and functions.[102]

Koala and Vicuna are some useful examples of fine-tuned models. Vicuna is an open-source bot[103] that was created by fine tuning LLaMA – 13B using 70000 conversations by users of Chat GPT downloaded from ShareGPT. To evaluate Vicuna's capability, GPT-4 was used as a judge[104] of the bot's responses in comparison to Bard, ChatGPT, Alpaca-13B and LLaMA-13B. The metrics were detail, helpfulness, relevance and accuracy. Vicuna's total score was 92% quality relative to ChatGPT. It was preferred over LLaMA and Alpaca in more than 90% of the questions and was rated as better or equal to ChatGPT in 45% of the questions.[105] Koala is another chatbot fine-tuned on LLaMA using freely available dialogue data from the internet. Based on human trials and feedback, it often exceeded the performance of Alpaca and was also able to match and even exceed the quality of ChatGPT in a large number of cases.[106]

That said, it is well recognized that current chatbot benchmarks are not trustworthy ways to accurately evaluate the performance levels of chatbots. Performance measurements are currently based on the similarity of the finetuning data to the benchmark dataset. When more targeted evaluations are conducted, the capabilities gap between existing open-source models and the closed ones remain, and the imitation models do not perform nearly as well.[107] Imitation models therefore seem to have a less extensive knowledge base as compared to their larger base models. It appears that they only learn to mimic the style (confident and well-structured answers) of the model's behavior on the specific task.

---

[100] Shirin Ghaffary, 'Why Meta is giving away its extremely powerful AI model,' Vox, 28 July 2023, https://www.vox.com/technology/2023/7/28/23809028/ai-artificial-intelligence-open-closed-meta-mark-zuckerberg-sam-altman-open-ai accessed on 26 October 2023.

[101] Anthropic, 'Anthropic's responsible scaling policy- version 1.0,' September 2023, https://www-files.anthropic.com/production/files/responsible-scaling-policy-1.0.pdf, page 7.

[102] Cameron Wolfe, 'Imitation models and the open- source LLM revolution,' Deep (Learning) Focus, 19 June 2023, https://cameronrwolfe.substack.com/p/imitation-models-and-the-open-source accessed on 26 October 2023.

[103] The training and hosting code for Vicuna is publicly available.

[104] Lianmin Zheng et al, 'Judging LLM- as-a-judge- with MT- Bench and Chatbot Arena,' arxiv, 2023, https://arxiv.org/pdf/2306.05685.pdf, page 4. Using GPT-4 as an automated evaluation framework was proposed in this paper because of its ability to produce consistent ranks and detailed assessments.

[105] The Vicuna Team, 'Vicuna: An open- source chatbot impressing GPT-4 with 90%* ChatGPT quality,' Large Model Systems Organization, 30 March 2023, https://lmsys.org/blog/2023-03-30-vicuna/ accessed on 26 October 2023.

[106] Xinyang Geng et al, 'Koala: A dialogue model for academic research,' Berkeley Artificial Intelligence Research, 3 April 2023, https://bair.berkeley.edu/blog/2023/04/03/koala/ accessed on 26 October 2023.

However, experiments show that an imitation model can be truly better if a larger and more comprehensive imitation dataset and a better base LLM are used. Orca is an example of a smaller model that has been developed through imitation of the reasoning process of large foundational models. Additionally, Orca was fine-tuned using a larger and more complex dataset and explanation tuning, which involves giving the imitation model explanations for the larger base model's responses. It has performed competitively against ChatGPT,[108] something that shows it is possible to genuinely improve model capabilities using some methods.

Through the strategies of data curation and fine-tuning discussed above, research and practice demonstrate that it is possible to produce equally powerful imitation models. But how long does it take to build such high-quality imitation models, and how costly is it? Methods like Low-Rank Adaptation (LoRA) and Quantized Low-Rank Adaptation (QLoRA) have allowed for efficient finetuning on small amounts of hardware.[109] LoRA makes it possible to fine-tune LLMs on a single GPU by allowing you to train fewer weights i.e., those specific to your use case. This is done by freezing the pretrained weights and introducing a smaller set of trainable parameters (matrices) into the model's layers. This is particularly impressive because it does not increase the time it takes for the model to make the next prediction once fed with input as other fine-tuning adapters do. Additionally, model quality does not recede.[110] QLoRA goes even further by reducing the average memory requirements of fine tuning a 65 billion parameter model from 780+ GB of GPU memory to -48 GB of GPU memory without lowering performance. This approach has made it possible to create the largest publicly available models fine-tunable on a single GPU.

The decline in resources required has led to a "significant shift in accessibility."[111] Now, more actors can fine tune, which means that these open-source models could be vulnerable to misuse. In addition to the small amount of compute now required, the ease and possibility of creating and running smaller but more capable models could result in the creation and reintroduction[112] of potentially dangerous versions of these models. Reasoning from this situation, we believe it is possible that if highly capable AI models are created and then released open source in future, they could be fine-tuned into equally powerful (and potentially misaligned) imitation models.

---

[107] Arnav Gudibande et al, 'The false promise of imitating proprietary LLMs,' arxiv, 2023, https://arxiv.org/pdf/2305.15717.pdf, pages 6-7.

[108] Subhabrata Mukherjee et al, 'Orca: Progressive Learning from Complex Explanation Traces of GPT-4,' arxiv, 2023, https://arxiv.org/pdf/2306.02707.pdf, pages 13- 14.

[109] Derrick Mwiti, 'How to fine-tune Llama 2 with LoRA,' Mldive, 23 September 2023, https://www.mldive.com/p/how-to-fine-tune-llama-2-with-lora accessed on 26 October 2023.

[110] Edward Hu et al, 'LoRA: Low- Rank Adaptation of Large Language Models,' arxiv, 2021, https://arxiv.org/pdf/2106.09685.pdf, page 4.

[111] Tim Dettmers et al, 'QLoRA: Efficient Finetuning of Quantized LLMs,' arxiv, 2023, https://arxiv.org/pdf/2305.14314.pdf, page 2.

### B. Will some leading AI developers continue releasing frontier models open source?

Top officials at Meta have been forthright that they believe openly sharing the code and weights of AI models is useful for humanity. Mark Zuckerberg, Meta's CEO and Founder, is on record that he "believes it's the smartest thing to do to spread the company's influence and ultimately move faster toward the future."[113] Yann LeCun, Meta's Chief AI Scientist, has argued that closed-ness is *considerably* more dangerous than open-ness.[114] In a statement on the release LLaMA, the company emphasized that open access makes it easier to identify and solve problems faster as a community.[115] In response to safety concerns, the Vice President of AI research at Meta said "I think these are valid concerns, but the only way to have conversations in a way that really helps us progress is by affording some level of transparency".[116] These stances are particularly noteworthy because Meta is one of the world's leading AI companies. Indeed, few would argue that their models are not part of the few very advanced models in existence. They also have the resources to continue being at the forefront of AI development.[117]

Meta is not isolated in thinking about models this way. Recently, several important AI developers and ecosystem actors including Dell, IBM, Sony and Intel coalesced to form an 'AI Alliance' to promote open source AI development.[118] To be fair, some companies developing LLMs and doing some fine tuning on LLaMA (for example Hugging Face) remain hesitant about releasing their models open-source primarily citing misuse concerns.[119] There is also a case to be made that open sourcing will inevitably get shelved

---

[112] Elizabeth Seger et al, 'Open- sourcing highly capable foundation models: An evaluation of risks, benefits and alternative methods for pursuing open- source objectives,' Center for the Governance of AI, 2023, https://cdn.governance.ai/Open-Sourcing_Highly_Capable_Foundation_Models_2023_GovAI.pdf, page 12.

[113] Cade Metz and Mike Isaac, 'In Battle Over A.I, Meta decides to give away its crown jewels,' The New York Times, 18 May 2023, https://www.nytimes.com/2023/05/18/technology/ai-meta-open-source.html accessed on 26 October 2023.

[114] Yann LeCun, Twitter post, Twitter, 18 May 2023, https://twitter.com/ylecun/status/1659172655663030272?lang=en.

[115] Meta, 'Meta and Microsoft introduce the next generation of Llama,' Meta Newsroom, 18 July 2023, https://about.fb.com/news/2023/07/llama-2/ accessed on 27 October 2023.

[116] Sharon Goldman, 'With a wave of new LLMs, open- source AI is having a moment and a red - hot debate,' VentureBeat, 10 April 2023, https://venturebeat.com/ai/with-a-wave-of-new-llms-open-source-ai-is-having-a-moment-and-a-red-hot-debate/ accessed on 27 October 2023.

[117] Wes Davis, 'Meta sets GPT-4 as the bar for its next AI model, says a new report', The Verge, 11 September 2023, https://www.theverge.com/2023/9/10/23867323/meta-new-ai-model-gpt-4-openai-chatbot-google-apple accessed on 27 October 2023.

[118] 'Meta and IBM launch 'AI Alliance' to promote open-source AI development' The Guardian, 5 December 2023, https://www.theguardian.com/technology/2023/dec/05/open-source-ai-meta-ibm#:~:text=The%20AI%20Alliance%20–%20led%20by,ideas%20and%20on%20open%20innovation%2C, accessed on 6 December 2023.

[119] Will Heaven, 'The open- source AI boom is built on Big Tech's handouts. How long will it last?' MIT Technology Review, 12 May 2023, https://www.technologyreview.com/2023/05/12/1072950/open-source-ai-google-openai-eleuther-meta/ accessed on 27 October 2023. Hugging Face's Chief Ethics Scientist Margaret

once the commercial concerns win out.[120] Whatever the case, we think there is ample evidence showing that the idea of open sourcing powerful models is certainly not on its knees.

### C. Do global south countries have actors with the capability and resources to fine tune open-source models into more risky models?

There is evidence that some global south countries are fairly competitive in the research and development of AI. For example, recent iterations of the Global AI Index[121] reveal that despite the United States and China maintaining their grip on the 1st and 2nd positions respectively,[122] Singapore has moved up significantly. In 2021, it rose from 10th to 6th place and in 2023 it rose to third place.[123] Per the 4th update of the Index, AI capacity in Singapore is ranked 1st on the metric of AI intensity and 3rd overall.[124] On the metric of talent, India is ranked second after the United States, with Singapore placed 4th.

Another ranking can be found in Stanford University's AI Report Index, which provides a measure of national AI capabilities by presenting a comparison of the number of significant models produced in different countries. In the 2023 Index, the US stood out with 16 significant machine learning systems. It was followed by the UK with 8, China with 3, Canada and Germany with 2. They were followed by France, India, Russia and Singapore—each with 1.[125] India was also recognized for its steadily increasing share of publications in AI Journals.[126] According to the 2022 Index, software developers in India contributed a large portion (24.2% relative to the EU and UK's 17.3% and the US' 14%) of GitHub AI projects[127] and were considered the top region with highest AI skill penetration rates.[128]

---

Mitchell favors 'responsible democratization' which resulted in the company restricting access to people who have legitimate reasons to access the models via an, approval process.

[120] Dylan Patel and Afzal Ahmad, 'Google "Will have no moat, and neither does OpenAI,"' Semianalysis, 4 May 2023, https://www.semianalysis.com/p/google-we-have-no-moat-and-neither accessed on 27 October 2023.

[121] Alex Mostrous, Joseph White, Serena Cesareo, 'The Global Artificial Intelligence Index,' Tortoise, 28 June 2023, https://www.tortoisemedia.com/2023/06/28/the-global-artificial-intelligence-index/ accessed on 27 October 2023.

[122] Lizka, 'The state of AI in different countries – an overview,' Effective Altruism Forum, 14 September 2023, https://forum.effectivealtruism.org/posts/Lb2TjSsjpqA8rQ7dP/the-state-of-ai-in-different-countries-an-overview#Where_frontier_AI_progress_is_happening_right_now accessed on 27 October 2023; Tortoise, 'The Global AI Index,' June 2023, https://www.tortoisemedia.com/intelligence/global-ai/; Stanford University Human-Centered Artificial Intelligence, 'Artificial Intelligence Index Report,' 2023, https://aiindex.stanford.edu/wp-content/uploads/2023/04/HAI_AI-Index-Report_2023.pdf.

[123] Alex Mostrous, Joseph White, Serena Cesareo, 'The Global Artificial Intelligence Index,' Tortoise, 28 June 2023, https://www.tortoisemedia.com/2023/06/28/the-global-artificial-intelligence-index/ accessed on 27 October 2023.

[124] Tortoise, 'The Global AI Index,' June 2023, https://www.tortoisemedia.com/intelligence/global-ai/. It is followed by Israel and Switzerland.

[125] Stanford University Human-Centered Artificial Intelligence, 'Artificial Intelligence Index Report,' 2023, https://aiindex.stanford.edu/wp-content/uploads/2023/04/HAI_AI-Index-Report_2023.pdf, page 51.

[126] Stanford University Human-Centered Artificial Intelligence, 'Artificial Intelligence Index Report,' 2023, https://aiindex.stanford.edu/wp-content/uploads/2023/04/HAI_AI-Index-Report_2023.pdf, page 34.

[127] Stanford University Human-Centered Artificial Intelligence, 'Artificial Intelligence Index Report,' 2023, https://aiindex.stanford.edu/wp-content/uploads/2023/04/HAI_AI-Index-Report_2023.pdf, page 67.

[128] Stanford University Human-Centered Artificial Intelligence, 'Artificial Intelligence Index Report,' 2023, https://aiindex.stanford.edu/wp-content/uploads/2023/04/HAI_AI-Index-Report_2023.pdf, page 182.

We believe that this assessment demonstrates 2 points: First, global south countries such as India and Singapore already have the kind of capability and resources to fine tune frontier models to equally capable (and potentially risky) imitation models. Second, serious investment in other global south countries' national capabilities could also allow them to compete on a global scale. While African countries still fall behind,[129] it remains likely that the few existing structures in some of those countries (particularly the ones that feature in the ranking: South Africa, Nigeria and Kenya) will see more investment as advanced systems are developed and made more widespread.

### D. How global south countries can reduce the chances that highly capable misaligned AI will be created through fine tuning within their jurisdictions

If open sourcing of frontier AI models continues, it could accelerate progress in the development of even more capable AI, allow for decentralized control of these systems and possibly enable accountability and transparency in the oversight of AI.[130] However, open sourcing also carries serious risks. This may happen through: (i) the unintended creation of smaller, powerful but harmful models as has been suggested, and (ii) the malicious use of these open-source foundational systems. So, what can the global south countries where such fine tuning could potentially happen do about this?

First, they can promulgate regulations or industry-recommended guidelines and standards that require developers involved in fine tuning the most advanced frontier AI models to conduct risk assessments and disclose to a government agency any apparent issues and steps taken to mitigate them. This may seem like an onerous responsibility to foist upon some developers but in many global south countries cybersecurity law already has analogous reporting requirements for entities with fairly small turnovers.[131] The risk assessments that these fine-tuning developers are expected to conduct do not have to be exactly similar to those which the large AI companies conduct. In recognition of their limited resources, perhaps the assessments can initially be less exacting and demanding.

Second, these countries could consider making it mandatory for models which have higher or comparable capabilities to existing foundational models ("comparably capable fine-tuned models") to have built-in monitoring capabilities and controlled release. This would be in line with suggestions that have been made for foundational models that are highly capable.[132] Furthermore, if the assessments conducted reveal severe risk or a high level of risk uncertainty, the developer of the comparably capable fine-tuned model should be required to obtain approval for release.

We concede that our proposals here will only work if the appropriate government bodies are well-resourced, efficient and robust, and this has never been a strong suit of most global south countries. In response to that point, we suggest that the countries which are likely to be places where comparably capable fine-tuned models are built (for example India or Singapore) have shown that they can operate government bodies in a way that meets the criteria above if a goal is deemed sufficiently important. Their assessment, monitoring and investigation of tax compliance is one example in line with our contention here.

---

[129] South Africa is 55th, Nigeria 60th and Kenya comes in last as 62nd.

[130] Elizabeth Seger et al, 'Open- sourcing highly capable foundation models: An evaluation of risks, benefits and alternative methods for pursuing open- source objectives,' Center for the Governance of AI, 2023, https://cdn.governance.ai/Open-Sourcing_Highly_Capable_Foundation_Models_2023_GovAI.pdf, pages 2,17-28.

[131] Consider for example Section 40 of Kenya's Computer Misuse and Cybercrimes Act (Act No. 5 of 2018).

[132] Markus Anderljung et al, 'Frontier AI regulation: Managing emerging risks to public safety,' arxiv, 2023, https://arxiv.org/pdf/2307.03718.pdf, page 26.

## 3. STRATEGIC REASONS WHY THESE COUNTRIES COULD HAVE INFLUENCE ON THE PATH TO HIGHLY CAPABLE MISALIGNED AI

### 3.1 In general, these countries can have an impact on the creation and use of multilateral rules and institutions to ensure safe development of highly capable AI

#### A. Global south countries have historically played a crucial role in the design of multilateral rules and institutions

The design of multilateral rules and institutions plays a pivotal role in the evolving landscape of international relations. By 'design,' we mean more than just a structural outline; we are referring to the foundational architecture that determines how a rule or institution works at multiple levels—from what is mandated to governance structures.[133] Design is the bedrock upon which efficacy rests. Without thoughtful design, a rule or institution can become a quagmire of inefficiency and ineffectiveness, often exacerbating the very problems it aims to solve.[134]

Global south countries have been instrumental in shaping the design of multilateral rules and institutions in the past, offering critical perspectives that have led to real results. For instance, the Non-Aligned Movement (NAM), consisting primarily of states from the global south, played a pivotal role in shaping the Nuclear Non-Proliferation Treaty (NPT) during its review conferences.[135] The NPT, originally constructed with a bias favoring the nuclear-armed states, was significantly democratized by the continuous and strategic efforts of NAM states. They actively engaged in diplomatic dialogues and negotiations, scrutinizing the inherent inequalities, and pushing for revisions and modifications to establish a more balanced power dynamic within the treaty framework. They emphasized the necessity of a transparent and equitable review process that could comprehensively reflect the concerns, aspirations, and interests of both nuclear and non-nuclear states.[136]

---

[133] Josep Colomer, 'Institutional design,' in Todd Landman and Neil Robinson (editors), The SAGE Handbook of Comparative Politics, Sage Publications, London, 2009, page 246; Erik Voeten, 'Making sense of the design of international institutions,' Volume 22, Annual Review of Political Science, 2019, https://www.annualreviews.org/doi/full/10.1146/annurev-polisci-041916-021108, page 146.

[134] Robert Keohane, After Hegemony: Cooperation and Discord in the World Political Economy, 1st edition, Princeton University Press, Princeton, 1984, page 59.

[135] International Atomic Energy Agency, '1995 review and extension conference of the parties to the treaty of Non-Proliferation of Nuclear weapon, ' June 1995, https://www.iaea.org/sites/default/files/publications/documents/infcircs/1995/infcirc474.pdf, page 4.

[136] Matthew Harries, 'Disarmament as politics: Lessons from the negotiation of NPT Article VI,' International Security, 2015, https://www.chathamhouse.org/sites/default/files/field/field_document/20150512DisarmamentPoliticsNPTHarries.pdf, page 6.

The common but differentiated responsibilities (CBDR) principle, integral to the United Nations Framework Convention on Climate Change (UNFCCC), also shows the global south's crucial role in the design of multilateral institutions.[137] The CBDR principle is important not merely as a rule but as an integral component in shaping the architecture and governance mechanisms of the UNFCCC.[138] It mandates a nuanced and balanced approach to international climate governance, allowing the disparities in national resources, developmental stages, and historical emissions of developing nations to be at the forefront of decision-making processes. Through the implementation of the CBDR principle, the institutional framework of the UNFCCC is designed to reflect the voices and concerns of global south countries, enabling these nations to actively participate and influence the global climate discourse.

Another salient example comes from the formation of the United Nations Conference on Trade and Development (UNCTAD).[139] Here the Group of 77, primarily consisting of countries from the global south, successfully advocated for a consensus-based decision-making process.[140] The Group's advocacy for a consensus-based decision-making process was not just a procedural choice—it was a fundamental redesign of institutional decision-making. By insisting on consensus, they ensured every nation, regardless of its size or economic influence, had an equal stake in decisions.[141] This design choice has made UNCTAD a more democratic platform, offering a substantive stage for the interests of developing nations.[142]

Through these case studies, it is evident that global south countries contributions extend far beyond surface-level adjustments. They have been central to the formation and reformation of the foundational architecture of numerous multilateral rules and institutions.

### B. Global south countries have historically played a crucial role in the use of multilateral rules and institutions

It also seems useful to remind ourselves that global south countries have historically influenced how multilateral rules and institutions are used. When discussing the "use" here, we are referring to their practical application and the mechanisms that ensure their mandates are effectively executed. This incorporates everything from the

---

[137] Malek Romdhane, 'What is the common but differentiated responsibilities and respective capabilities (CBDR-RC) principle?' ClimaTalk Contributor, 21 July 2021, https://climatalk.org/2021/07/12/what-is-the-cbdr-rc-principle/ accessed on 28 August 2023.

[138] Shayne Kavanagh et al, 'A framework for a financial sustainability index,' Lincoln Institute of Land Policy, April 2017, https://www.lincolninst.edu/publications/working-papers/framework-financial-sustainability-index, page 29.

[139] United National Conference on Trade and Development, 'History,' https://unctad.org/about/history accessed on 28 August 2023.

[140] Karl Sauvant, 'The early days of the Group of 77,' Volume 51 Issue 1, United Nations Chronicles, 2014, https://www.un.org/en/chronicle/article/early-days-group-77 accessed on 28 August 2023.

[141] Shrirang Shukla, 'From GATT to WTO and beyond,' UNU World Institute for Development Economics Research, 2000, https://www.wider.unu.edu/sites/default/files/wp195.pdf, pages 7-9.

[142] Marc Williams, 'The Group of 77 in UNCTAD: Anatomy of a third world coalition,' London School of Economics and Political Science University of London, Ph.D Thesis, 1987, https://core.ac.uk/download/pdf/4187564.pdf, page 164.

implementation of formulated guidelines to the real-world enforcement of international standards.[143]

To illustrate the scope of influence, one can turn to historical precedents such as South Africa's pivotal role in nuclear development advocacy.[144] Having dismantled its own nuclear arsenal, South Africa became an influential voice in the global nuclear disarmament movement and played a pivotal role in shaping the discussions and resolutions at the International Atomic Energy Agency (IAEA).[145] Its experience in both developing and relinquishing nuclear capabilities made it uniquely qualified to offer insights on responsible nuclear policy, providing a roadmap for other countries to follow.

Global south countries are especially instrumental in influencing the use of multilateral rules and institutions when they buy into certain narratives. In the part that immediately succeeds this one, we discuss in depth the relevance of narratives and how they shape policy especially at a global level. We will show that through the narratives they adopt, global south countries have profoundly shaped the dynamics of some important multilateral rules and institutions.

### C. The narratives that global south countries adopt are particularly consequential to the design and use of multilateral rules and institutions

In public policy discourse, it is largely accepted that narratives are "meaning-making tools." The understanding is that stories are constitutive of human existence, which makes it difficult to imagine communicating without them.[146]

Narratives are considered "strategic constructions of a policy reality promoted by policy actors seeking to win in public policy battles."[147] They are constructed on the basis that: (i) it is the perceptions and interpretations of reality that matter for public policy and (ii) narratives contain identifiable and measurable elements that can be categorized as generalizable. These elements include: the setting in which the problem a policy narrative is addressing exists, the distinct characters portrayed in the narrative, the interaction between these characters and their contexts, and the proposals for ways to solve and approach the challenges the communities' face.[148]

---

[143] Erik Voeten, 'Making sense of the design of international institutions,' Volume 22, Annual Review of Political Science, 2019, https://www.annualreviews.org/doi/full/10.1146/annurev-polisci-041916-021108, pages 147-164; Thomas Gehring and Sebastian Oberthur, 'Institutional interaction in global environmental governance: The case of the Cartagena Protocol and the World Trade Institution,' Volume 6 Issue 2, Global Environmental Politics, 2006, https://direct.mit.edu/glep/article/6/2/1/14352/Institutional-Interaction-in-Global-Environmental, pages 3-9.

[144] World101, 'South Africa: Why countries acquire and abandon nuclear bombs,' Council on Foreign Relations, 27 July 2023, https://world101.cfr.org/global-era-issues/nuclear-proliferation/south-africa-why-countries-acquire-and-abandon-nuclear accessed on 17 August 2023.

[145] World101, 'South Africa: Why countries acquire and abandon nuclear bombs,' Council on Foreign Relations, 27 July 2023, https://world101.cfr.org/global-era-issues/nuclear-proliferation/south-africa-why-countries-acquire-and-abandon-nuclear accessed on 17 August 2023.

[146] Michael Jones, Elizabeth Shanahan, and Mark McBeth, 'Introducing the Narrative Policy Framework,' in Michael Jones, Elizabeth Shanahan, and Mark McBeth (editors), The Science of Stories: Applications of the Narrative Policy Framework in Public Policy Analysis, Palgrave Macmillan, 2014, page 9.

[147] Michael Jones, Elizabeth Shanahan, and Mark McBeth, 'A brief introduction to the Narrative Policy Framework,' in Michael Jones, Elizabeth Shanahan, and Mark McBeth (editors), Narratives and the Policy Press: Applications of the Narrative Policy Framework, Montana State University Library, 2022, page 2.

Throughout human history, various narratives can be identified to have played a major role in effecting political and economic change.[149] In what follows, we discuss some of the narratives that global south countries adopted in international environmental governance and international governance of intellectual property rights. We also discuss the impacts of these narratives on the situations in question.

### (i) How narratives adopted by global south countries impacted international governance to combat climate change

**The 'low carbon transition will hinder economic growth' narrative**

At some point in time, a number of developing countries[150] developed and adopted a narrative that an energy transition to reduce greenhouse gas (GHG) emissions (the 'low carbon transition') would hinder their overall economic growth. The connection is not very surprising. Historically, economic development has been associated with burning fossil fuels, which are now understood to be a major source of GHG emissions.[151] The narrative became powerful because these fossil fuels remain the most reliable source of energy and because it is widely recognized that Least Developed Countries (LDCs) bear the least historical responsibility for climate change.[152]

**India as a case study**

When COP 21 was convened in Paris, it was intended to result in what was understood to be the 'most significant international agreement yet to reduce global greenhouse gas emissions and slow the effect of climate change'.[153] While there was unity on the overall goal, the question of how the developed and developing nations would divide responsibility was one of the most disputatious. At the time, India had just come out as a central figure in defining the agreement for two reasons: (i) it was (and still is) ranked as the world's third largest carbon emitter[154] and (ii) it has deemed it necessary to continue to be structurally dependent on coal to alleviate poverty, increase their population's access to modern energy and industrialize.[155] Notably, India positioned herself as a leader of the developing world in the negotiations that followed.[156]

---

[148] Michael Jones, 'Advancing the narrative policy framework? The musings of a potentially unreliable narrator,' Volume 46 Issue 4, Policy Studies Journal, 2018, https://onlinelibrary.wiley.com/doi/abs/10.1111/psj.12296, page 727.

[149] Michael Jones, 'Communicating Climate Change: Are Stories Better than "Just the Facts"?,' Volume 43 Issue 14, Policy Studies Journal, 2014, https://onlinelibrary.wiley.com/doi/abs/10.1111/psj.12072, page 645.

[150] Amar Bhattacharya, Homi Kharas and John McArthur, 'Developing countries are key to climate action,' Brookings, 3 March 2023, https://www.brookings.edu/articles/developing-countries-are-key-to-climate-action/ accessed on 28 October 2023.

[151] Tim Gore, Mira Alestig and Anna Ratcliff, 'Confronting carbon inequality: Putting climate justice at the heart of the COVID-19 recovery,' Oxfam Media Briefing, 21 September 2020, https://oxfamilibrary.openrepository.com/bitstream/handle/10546/621052/mb-confronting-carbon-inequality-210920-en.pdf accessed on 28 October 2023.

[152] United Nations Conference on Trade and Development, 'The Least Developed Countries report 2022: The low-carbon transition and its daunting implications for structural transformation,' 3 November 2022, https://unctad.org/ldc2022 accessed on 28 October 2023. In 2019, LDCs were estimated to have accounted for about 1.1% of the total world Carbon dioxide emissions from fossil- fuel combustion and industrial processes.

[153] Justin Worland, 'What to know about the Paris Climate Change Conference,' Time, 29 November 2015, https://time.com/4123568/paris-climate-conference-preview/ accessed on 28 October 2023.

[154] Jocelyn Timperley, 'The Carbon Brief Profile: India,' Carbon Brief Clear on Climate, 14 March 2019, https://www.carbonbrief.org/the-carbon-brief-profile-

At the plenary session, India's Prime Minister asserted that climate justice demands that developing countries should have enough space to grow with the carbon space still available. He also insisted that advanced nations should set ambitious targets to reduce their emissions not only because of their historical responsibility but also because "they have the most room to make these cuts and make the strongest impact".[157] He defended this principle of differentiation (that developed countries should take up more stringent responsibilities) and insisted that it ought to be reflected across all sections and provisions of the agreement under consideration. Of course, his position here reflects a narrative that has enjoyed widespread popularity in global south countries.[158]

**Consequence of narrative adoption**

In the final version of the Paris Agreement adopted,[159] the principle of equity and common but differentiated responsibilities was explicitly recognized and restated as one of the key goals. The principle is specifically highlighted in Article 4, which provides that developed countries *should* take the lead by undertaking economy-wide absolute emission reduction targets.[160] On the other hand, the agreement only *encourages* developing countries to move towards the goals set out. There is no doubt that the forcefulness with which India (and other global south countries in support) pushed for this made a big difference to the resulting agreement.

---

india/#:~:text=India%20is%20the%20world's%20third,after%20China%20and%20the%20US accessed on 28 October 2023.

[155] Samir Saran, 'Deconstructing Indian Leadership on Climate Change,' myGov, 7 December 2015, https://blog.mygov.in/editorial/editorial-by-samir-saran/ accessed on 28 October 2023. 'Coal is still a necessity for multiple lifeline initiatives of the country to lift millions out of poverty.'

[156] John Vidal, 'India pushes rich countries to boost their climate pledges at Paris,' The Guardian, 2 December 2015, https://www.theguardian.com/environment/2015/dec/02/india-takes-leading-role-for-global-south-nations-in-climate-talks accessed on 28 October 2023.

[157] Narendra Modi, 'Prosperous have strong carbon footprint, world billions at the bottom of development ladder seek space to grow: PM at COP 21 Plenary,' 30 November 2023, https://www.narendramodi.in/pm-modi-addresses-the-plenary-session-at-cop-21-summit-in-paris-385298 accessed on 28 October 2023.

[158] Joydeep Gupta, 'Explained: The negotiating blocs that will steer COP26,' China Dialogue, 1 November 2012, https://chinadialogue.net/en/climate/explained-the-negotiating-blocs-that-will-steer-cop26/ on 22 August 2023; Sumudu Atapattu and Carmen Gonzales, 'The north-south divide in International Environmental Law: Framing the issues,' in Shawkat Alam et al (editors), International Environmental Law and the Global South, 1st edition, Cambridge University Press, New York, 2015, page10.

[159] Paris Agreement on Climate Change, 3156 UNTS, 12 December 2015, https://unfccc.int/sites/default/files/resource/parisagreement_publication.pdf.

[160] Article 4 Paragraph 4, Paris Agreement on Climate Change, 3156 UNTS, 12 December 2015, https://unfccc.int/sites/default/files/resource/parisagreement_publication.pdf.

### (ii) How narratives adopted by global south countries impacted international governance of intellectual property rights

**The 'addressing public health concerns is more important than protecting intellectual property rights' narrative**

Following the Uruguay Round negotiations (1986-1994), the Agreement on Trade Related Aspects of Intellectual Property Rights (TRIPS Agreement) was adopted to provide restrictive and extensive patent protection on pharmaceutical products.[161] All countries, for instance, were required to provide patent protection for all technology for a minimum period of 20 years.[162] Furthermore, developing countries were mandated to amend their laws to become TRIPS compliant.[163]

Years later, developing countries were facing widespread public health challenges. In particular, HIV/ AIDS was ravaging the population of countries like South Africa[164] and Brazil.[165] With the lead of South Africa, the narrative that addressing public health concerns is more important than protecting intellectual property rights took root among developing countries.[166]

**Consequences of narrative adoption**

Motivated by this narrative, the African Group[167] made a proposal to members of TRIPS Council to hold a special meeting for discussions on the relationship between access to medicine and intellectual property rights (IPRs). This session was intended to initiate the process of the interpretation and application of the TRIPS Agreement in a way that did not undermine the implementation of public health policies.[168] It had the ultimate aim of confirming the right of World Trade Organization (WTO) members to use the TRIPS Agreement's public health safeguards.

---

[161] This is in comparison to prior agreements on the protection of intellectual property rights, for example the Paris Agreement. See Divya Murthy, 'The Future of compulsory licensing: Deciphering the Doha Declaration on the TRIPS Agreement and Public Health,' Volume 17 Issue 6, American University International Law Review, 2002, https://digitalcommons.wcl.american.edu/cgi/viewcontent.cgi?referer=&httpsredir=1&article=1232&context=auilr, pages 1308-1310; James Gathii, 'The legal status of the Doha Declaration on TRIPS and Public Health under the Vienna Convention on the Law of Treaties,' Volume 15 Issue 2, Harvard Journal of Law and Technology, 2002, https://lawecommons.luc.edu/cgi/viewcontent.cgi?referer=&httpsredir=1&article=1419&context=facpubs page 294.

[162] Article 33, Agreement on Trade-Related Aspects of Intellectual Property Rights, 1869 U.N.T.S. 299, 15 April 1994, https://www.wto.org/english/docs_e/legal_e/trips_e.htm#art5.

[163] International Institute for Sustainable Development, 'TRIPS and Public Health,' IISD Trade and Development Brief Number 9, 2003, https://www.iisd.org/system/files/publications/investment_sdc_dec_2003_9.pdf, page 2.

[164] Mark Horton, 'HIV/AIDS in South Africa,' in Michael Nowak and Luca Ricci (editors), Post-Apartheid South Africa: The First Ten Years, The International Monetary Fund, 2005, page 113- 114; Didier Fassin and Helen Schneider, 'The politics of AIDS in South Africa: Beyond the controversies,' Volume 326, British Medical Journal, 2003, https://www.ncbi.nlm.nih.gov/pmc/articles/PMC1125376/pdf/495.pdf, page 495.

[165] Vera Zolotaryova, 'Are we there yet? Taking TRIPS to Brazil and expanding access to HIV/AIDS medication,' Volume 33 Issue 3, Brooklyn Journal of International Law, 2008, https://brooklynworks.brooklaw.edu/cgi/viewcontent.cgi?article=1209&context=bjil, page 1110.

[166] James Gathii, 'The legal status of the Doha Declaration on TRIPS and Public Health under the Vienna Convention on the Law of Treaties,' Volume 15 Issue 2, Harvard Journal of Law and Technology, 2002, https://lawecommons.luc.edu/cgi/viewcontent.cgi?referer=&httpsredir=1&article=1419&context=facpubs page 296.

[167] World Trade Organization, 'Groups in the Negotiation,' 12 April 2021, https://www.wto.org/english/tratop_e/dda_e/negotiating_groups_e.htm on 28 October 2023. This group consists of 44 countries together with Barbados, Bolivia, Brazil, Cuba, Dominican Republic, Ecuador, Honduras, India, Indonesia, Jamaica, Pakistan, Paraguay, Philippines, Peru, Sri Lanka, Thailand, and Venezuela.

In the Doha Ministerial Declaration of November 2001, the WTO members stressed the importance of implementing and interpreting the TRIPS Agreement in a manner supportive of public health, by promoting both access to existing medicines and research and development into new medicines.[169] To operationalize this position, these governments also adopted a separate declaration (Declaration on TRIPS and Public Health[170]) that recognized the need for international action to address the gravity of public health problems afflicting developing and least developed countries, highlighting HIV/AIDS, tuberculosis and malaria in specific.[171] This permitted certain flexibilities such as compulsory licensing under the TRIPS agreement—which allows governments to issue these licenses to allow other companies to make a patented product or use a patented process under license without the owner's consent under specific circumstances that ensure the protection of the legitimate interests of the patent holder.[172]

The Declaration also confirmed that WTO members were allowed to determine grounds for issuing compulsory licenses beyond emergency situations. These flexibilities acknowledged the inadequate pharmaceutical manufacturing capacities of certain WTO members and thus made it easier for poorer countries to import cheaper generics made under compulsory licensing. The declaration was a boon for the developing countries that had the capacity to manufacture generic drugs. For the countries with insufficient manufacturing capacity, relief came a little later. The WTO General Council Decision of 30 August 2003 allowed such countries to import generics produced under compulsory license subject to specific administrative requirements.[173]

---

[168] Submission by the African Group, Barbados, Bolivia, Brazil, Cuba, Dominican Republic, Ecuador, Honduras, India, Indonesia, Jamaica, Pakistan, Paraguay, Philippines, Peru, Sri Lanka, Thailand, and Venezuela, 'TRIPS and Public Health,' Council for Trade - Related Aspects of Intellectual Property Rights, IP/C/W/296, 29 June 2001, https://docs.wto.org/dol2fe/Pages/SS/directdoc.aspx?filename=Q:/IP/C/W296.pdf&Open=True.

[169] Doha WTO Ministerial Declaration, 14 November 2001, WT/MIN (01)/DEC/1, https://www.wto.org/english/thewto_e/minist_e/min01_e/mindecl_e.htm#trips, paragraph 17.

[170] Declaration on the TRIPS Agreement and Public Health, WT/MIN(01)/DEC/2, 14 November 2001, https://www.wto.org/english/thewto_e/minist_e/min01_e/mindecl_trips_e.htm.

[171] Declaration on the TRIPS Agreement and Public Health, WT/MIN(01)/DEC/2, 14 November 2001, https://www.wto.org/english/thewto_e/minist_e/min01_e/mindecl_trips_e.htm, paragraph 1.

[172] Declaration on the TRIPS Agreement and Public Health, WT/MIN(01)/DEC/2, 14 November 2001, https://www.wto.org/english/thewto_e/minist_e/min01_e/mindecl_trips_e.htm, paragraph 5(c).

[173] World Trade Organization, 'Members OK amendment to make health flexibility permanent,' World Trade Organization Press Releases 2005, 6 December 2005, https://www.wto.org/english/news_e/pres05_e/pr426_e.htm accessed on 28 October 2023. These requirements were: implementing measures to prevent re-exportation of drugs meant for developing countries to more lucrative markets, provision of notification of specific drugs to be imported and the exact quantity required and the various physical specifications for the medication that was meant to be consumed in these countries.

### (iii) What if global south countries adopt the narrative that the existence of highly capable AI will be crucial for their development?

It is usually assumed[174] that the terms and conditions for the access and use of new technologies are beyond the control of African countries. Yet it is possible that some narratives they adopt could impact how humanity goes about regulating the development of such technologies.

Consider a narrative one that positions highly capable AI as a critical tool for development. Already, some believe that the world is heading briskly towards transformative AI, which is understood to be 'AI that precipitates a transition comparable to or more significant than the agricultural or industrial revolution'.[175] This could happen if future AI systems have capabilities that fundamentally transform economies. It could also occur if AI systems prove capable of making significant contributions to science and engineering, and therefore cause a dramatic and possibly unexpected acceleration in the development of transformative technology. Given the demonstrated potential of the most advanced AI models available today, it would be understandable—even if not entirely accurate—for some to claim that highly capable AI will drive growth and development.

This is already a narrative that has begun gaining some traction in the conversation surrounding AI in global south countries. For example, a comprehensive report by Access Partnership in collaboration with University of Pretoria,[176] lays out a detailed analysis of the core sectors that AI could have long-term economic impact in. They include agriculture, healthcare, government/public services, financial services and education. The report goes further to cite a study that projects AI may double the GDP growth rate of countries like the US and China by 2035. Given this promise, the authors of the report claim that African countries should focus on adopting AI because even a fraction of such growth would transform their economies and help them reduce poverty drastically.

AI has also been cited as "a strategic priority for African countries" that should be leveraged and adopted to solve some of the continent's most pressing challenges. Evidence that the message is finding listening ears can be found in some of the steps being taken by African governments[177] as they explore the ways new technologies can be integrated and leveraged. Some steps taken include:[178]

---

[174] Nathaniel Allen, 'The promises and perils of Africa's digital revolution,' Brookings, 11 March 2021, https://www.brookings.edu/articles/the-promises-and-perils-of-africas-digital-revolution/ accessed on 28 October 2023.

[175] Holden Karnofsky, 'Some background on our views regarding advanced artificial intelligence,' Open Philanthropy, 6 May 2016, https://www.openphilanthropy.org/research/some-background-on-our-views-regarding-advanced-artificial-intelligence/ accessed on 28 October 2023.

[176] Access Partnership and University of Pretoria, 'Artificial intelligence for Africa: An opportunity for growth, development, and democratization,' November 2018, Artificial Intelligence for Africa: An Opportunity for Growth, Development, and Democratization, pages 8-12.

[177] Jake Effoduh, '7 Ways that African states are Legitimizing Artificial Intelligence,' openAIR, 20 October 2020, https://openair.africa/7-ways-that-african-states-are-legitimizing-artificial-intelligence/ accessed on 28 October 2023.

[178] Stefano Sedola, Andrea Pescino and Tira Greene, 'Blueprint: Artificial Intelligence for Africa,' Smart Africa, 2021, https://www.bmz-digital.global/wp-content/uploads/2022/08/70029-eng_ai-for-africa-blueprint.pdf, pages 24-25.

- The embrace of National AI Strategies as a tool of communication: "we understand the importance of AI to the world, and we have a plan"
- The creation of AI agencies and task forces
- The push for strategic partnerships to encourage local adoption of AI

It is plausible to argue that if governments are able to recognize the primary role of AI as a 'new factor of production',[179] then they are more likely to be actively involved in realizing the potential of AI for their nations. As a result of this, the narrative is likely to favor the implementation of AI-friendly regulation and initiatives.

The popularization of this narrative could result in these countries' supporting less strenuous international rules to allow for faster AI development. These countries could also create a conducive local regulatory environment for AI labs and companies to train their frontier models. For instance, this could mean relaxing the enforcement of the requirements surrounding data collection, storage and privacy of personal data that would be useful in training these highly capable systems. They could also endorse lax labor laws that allow for parts of AI development such as data annotation to continue to be done cheaply by global south workers. Effectively, such a narrative would make it easier for misaligned highly capable AI to be developed and released for use.

### 3.2 What could be done to increase the chances that global south countries will play a positive role in the use of multilateral rules and institutions to ensure safe development of AI?

First, we think it is important to proactively include global south countries in any multilateral conversations about AI governance. If these countries have to push and pull in order to get a voice, it is likely that this will result in deep suspicion, and we may well end up with weak global buy-in for any proposals. The most skeptical may suggest that this would be irrelevant if none of these countries can build highly capable AI. As we have shown in part 2 of the paper, we should be more circumspect about the presumption that actors in countries like India will not have the ability to build highly capable AI in future. Yet even if we accept that premise, it is worth noting that the side-lining of global south countries may in fact generate a surge of dissatisfaction within influential communities in leading global north countries. The idea of locking out countries where so many people will have to experience the impacts of highly capable AI will just not sit very well with many citizens of global north countries (some which, let us recall, have large immigrant communities from global south countries).

Second, we think it is important that efforts are made to ensure the narrative that highly capable AI will be desirable for global south development is not accepted without nuance. Narratives develop based on facts and sometimes fiction. It is critical for important actors in the decision-making of global south governments to be well informed about the full range of capabilities that AI may have. This means a comprehensive understanding of the risks posed by highly capable misaligned AI and what would be required to ensure that highly capable AI is safe enough to be developed and released for use. One practical way to pursue this goal could be through organizing training conferences where key actors are guided through the big questions.

---

[179] Stefano Sedola, Andrea Pescino and Tira Greene, 'Blueprint: Artificial Intelligence for Africa,' Smart Africa, 2021, https://www.bmz-digital.global/wp-content/uploads/2022/08/70029-eng_ai-for-africa-blueprint.pdf, pages 27- 29.

## 4. CONCLUSION & WAY FORWARD

In this paper we have shown how the path to developing highly capable misaligned AI may in fact meander through a global south country. We have also staked out some rough perspectives about how these countries can use their roles in the AI development and global governance ecosystem to reduce the chances that highly capable misaligned AI emerges. One argument we have not explored at length has to do with the possibility that highly capable AI will only emerge if models are trained to learn the world's physical environment.[180] Although we will not canvass that assertion in detail, if the argument ends up being correct then we foresee some circumstances in which some global south countries will be important. In particular, we think this will be the case where such models need to be trained to learn these countries' physical environments.

As a natural next step off our main argument, we think it would be useful and important for some skilled people to work on the posture that global south countries should take towards the development of highly capable AI. The term "work" here is used consciously, to capture a wide range of possible activities. That said, we think research and advocacy are the most helpful paths forward. With regards to research, we especially think that there is room for bespoke work on why global south countries like India should be concerned about the safety implications of high capability AI. There is also room for further work on the technical and strategic arguments we have discussed in this paper, others that we have touched on and more in-depth work on the specific ways in which these countries can ensure safe AI development.

With regards to advocacy efforts, we think it would be helpful for some people to work on convincing global south policy makers that highly capable AI poses a significant risk to human wellbeing and these countries should therefore put their weight behind governance strategies that aim to address the possibility of global catastrophe caused by highly capable AI. The exact ways in which they can do so would ideally be informed by the research that is carried out. In conclusion, we wish to pre-emptively clarify that our research should not be taken to imply that we believe preventing the development of highly capable AI that could cause a global catastrophe is obviously the most significant problem anyone could be working on. Instead, our position is that it is one of several important concerns that humanity needs to address in order to secure the wellbeing of current and future generations.

---

[180] Pei Wang and Patrick Hammer, 'Perception from and AGI perspective,' 11th International Conference- AGI 2018, Prague, 22-25 August 2018, https://cis.temple.edu/~pwang/Publication/perception.pdf, pages 1-2; Matthew Sadiku et al, 'Artificial general intelligence: A primer,' Volume 7 Issue 6, International Journal of Trend in Research and Development, http://www.ijtrd.com/papers/IJTRD22369.pdf, page 7; Federico Berruti, Pieter Nel and Rob Whiteman, 'An executive primer on artificial general intelligence,' McKinsey and Company, 19 April 2020 https://www.mckinsey.com/capabilities/operations/our-insights/an-executive-primer-on-artificial-general-intelligence accessed on 27 August 2023.